\documentclass[12pt]{article}
\usepackage{epsfig}

\def\a{\alpha}
\def\b{\beta}

\def\d{\delta}
\def\e{\epsilon}

\def\g{\gamma}
\def\h{\eta}

\def\j{\psi}
\def\k{\kappa}
\def\l{\lambda}
\def\m{\mu}
\def\n{\nu}

\def\r{\rho}
\def\s{\sigma}
\def\t{\tau}

\def\z{\zeta}
\def\D{\Delta}

\def\Q{\Theta}

\def\ve{\varepsilon}

\def\dg{\dagger}                                     
\def\wt#1{\widetilde{#1}}                    
\def\VEV#1{\left\langle #1\right\rangle}        
\def\beq{\begin{equation}}
\def\eeq{\end{equation}}

\def\bea{\begin{eqnarray}}
\def\eea{\end{eqnarray}}
\def\NO{\nonumber}
\def\Bar#1{\overline{#1}}

\def\pl#1#2#3{Phys.~Lett.~{\bf B {#1}} ({#2}) #3}
\def\np#1#2#3{Nucl.~Phys.~{\bf B {#1}} ({#2}) #3}
\def\prl#1#2#3{Phys.~Rev.~Lett.~{\bf #1} ({#2}) #3}
\def\pr#1#2#3{Phys.~Rev.~{\bf D {#1}} ({#2}) #3}

\def\ap#1#2#3{Ann.~of Phys.~{\bf {#1}} ({#2}) #3}

\begin{document}
\date{}

\title{ 
{\normalsize
\mbox{ }\hfill
\begin{minipage}{3cm}   
DESY 01-115\\
ZU-TH-32/01
\end{minipage}}\\
\vspace{2cm}
\bf CP Violation, Neutrino Mixing\\ and the Baryon Asymmetry}
\author{W.~Buchm\"uller\\
{\it Deutsches Elektronen-Synchrotron DESY, 22603 Hamburg, Germany}\\[7ex]
D.~Wyler\\
{\it Institut f\"ur Theoretische Physik, Universit\"at Z\"urich,
Z\"urich, Switzerland}
}
\maketitle

\thispagestyle{empty}

\vspace{2cm}
\begin{abstract}
\noindent
We study neutrino masses and mixings based on the simplest SO(10) mass
relations and the seesaw mechanism. We find that the requirement of
large neutrino mixings determines the relative magnitude of the heavy Majorana
neutrino masses in terms of the known quark mass hierarchy. This leads
to specific predictions for the structure of the neutrino mixing matrix, 
the light neutrino masses, CP violation in neutrino 
oscillations, neutrinoless double $\b$-decay and  the baryon asymmetry. 
\end{abstract}

\newpage

Recent results from the Sudbury Neutrino Observatory \cite{sno} and from
the Super-Kamiokande experiment \cite{sk01} provide further evidence for the
neutrino oscillation hypothesis as the solution of the solar neutrino
problem. Neutrino oscillations can also account for the atmospheric neutrino 
anomaly \cite{sk98,k2k}, and a consistent picture is obtained with
just three neutrinos, $\nu_e$, $\nu_\mu$ and $\nu_\tau$,  undergoing
`nearest neighbour' oscillations,  $\nu_e \leftrightarrow \nu_\mu$ and
$\nu_\mu \leftrightarrow \nu_\tau$.

The experimental results on the  $\nu_e$ deficit in the solar neutrino flux 
favour the LMA or LOW solutions \cite{bax98} of of the MSW conversion with 
large mixing angle. A large mixing also fits the atmospheric neutrino oscillations.
As a result, the leptonic mixing matrix
$U_{\alpha i}$ seems to be very different from the familiar CKM quark 
mixing matrix 
$V_{\alpha i}$. One finds $|U_{\alpha i}| = {\cal O}(1)$ for all
elements, except for  $|U_{e3}| < 0.16$ \cite{ft01}. Furthermore, a
possible hierarchy among the neutrino masses $m_i$ has to be much weaker
than the known hierarchy of quarks and charged leptons.

The seesaw mechanism naturally explains the smallness of light Majorana 
neutrino masses $m$ by the largeness of right-handed neutrino masses $M$ 
\cite{seesaw},  
\beq
m_{\nu}\simeq - m_D{1\over M}m_D^T \;, \label{seesaw}
\eeq
where $m_D$ is the Dirac neutrino mass matrix. In unified theories one
expects $m_D$ to be related to the quark and charged lepton mass
matrices. Since they have a large hierarchy, the almost non-hierarchical
structure of the leptonic mixing matrix is quite surprising.

The simplest grand unified theory (GUT) which unifies one generation of
quarks and leptons including the right-handed neutrino in a single 
irreducible representation is based on the gauge group SO(10) \cite{gfm75}.
In the following we shall demonstrate that, given the known properties 
of the up-quark mass matrix, the puzzle of large neutrino mixings can be 
resolved in SO(10) theories provided the heavy neutrino masses also obey 
a specific hierarchy. We then explore the consequences for several 
observables in neutrino
physics including the cosmological baryon asymmetry.
The role of the heavy neutrino mass hie\-rarchy for the light neutrino 
mixings has already been discussed in different contexts \cite{smi93,afm00,
falc,allbarr}.

In SO(10) theories
quark and lepton masses are obtained by coupling the fermion  multiplet
${\bf 16}=(q_L,u_R^c,e_R^c,d_R^c,l_L,\n_R)$ to the Higgs multiplets
$H_1({\bf 10})$, $H_2({\bf 10})$ and $\Phi({\bf 126})$,
\beq  
{\cal L} = h_{uij} {\bf 16}_i {\bf 16}_j H_1({\bf 10})
          +h_{dij} {\bf 16}_i {\bf 16}_j H_2({\bf 10})
          +h_{Nij} {\bf 16}_i {\bf 16}_j \Phi({\bf 126})\;.
\eeq
Here we have assumed that the two Higgs doublets of the standard model
are contained in $H_1$ and $H_2$, respectively\footnote{Note, that this is
unavoidable in models with SO(10) breaking by orbifold 
compactification \cite{abc01}.}. The corresponding Yukawa couplings are
\bea
{\cal L}_m 
&=& h_{uij}\Bar{q}_{Li}u_{Rj}H_1+h_{dij}\Bar{q}_{Li}d_{Rj}H_2\NO\\
&& + h_{\n ij}\Bar{l}_{Li}\n_{Rj}H_1+h_{eij}\Bar{l}_{Li}e_{Rj}H_2  
 + {1\over 2} h_{Nij}\Bar{\n}^c_{Ri}\n_{Rj}\Phi + h.c.\;.
\eea
The quark and lepton mass matrices
$m_u = h_u v_1$, $m_d = h_d v_2$, $m_D = h_\n v_1$ and $m_e = h_e v_2$,
with $v_1 = \VEV H_1$ and $v_2 = \VEV H_2$, satisfy the relations
\beq\label{muni}
m_u = m_D  \;, \quad m_d = m_e \;.
\eeq
Note, that all mass matrices are symmetric. The incorrect relation
$m_s=m_\mu$ can be corrected by contributions from higher dimensional
Higgs representations \cite{gj79}. The Majorana mass matrix
$M = h_N \VEV \Phi$ is a priori independent of $m_u$ and $m_d$. 

From the phenomenology of weak decays we know that the quark matrices have
approximately the form (see, e.g. \cite{fx00,rr01})
\beq\label{qmass}
m_{u,d} \propto \left(\begin{array}{ccc}
    0  & \e^3 e^{i\phi}  & 0 \\
    \e^3 e^{i\phi}  & \; \r\e^2 \; & \h\e^2 \\
    0  &  \h\e^2  & e^{i\j} 
    \end{array}\right) \;.
\eeq
Here $\e \ll 1$ is the parameter which determines the flavour
mixing, and 
\beq
\r = |\r| e^{i\a}\;, \quad \h = |\h| e^{i\b}\;,
\eeq
are complex parameters ${\cal O}(1)$.
We have chosen a `hierarchical' basis, where off-diagonal
matrix elements are small compared to the product of the corresponding
eigenvalues, $|m_{ij}|^2 \leq {\cal O}(|m_i m_j|)$. 
In contrast to the usual assumption of hermitian mass 
matrices \cite{fx00,rr01}, 
SO(10) invariance dictates the matrices to be symmetric.
All parameters may take different values
for up- and down- quarks. Typical choices for $\e$ are $\e_u = 0.07$, 
$\e_d = 0.21$ \cite{rr01}. The agreement with data can be improved by
adding in the 1-3 element a term ${\cal O}(\e^4)$ \cite{beg00,rrx01}
which, however, is not important for our analysis.

Three of the four phases in the quark mass matrix (\ref{qmass}) can be
absorbed into a phase matrix $P$,
\beq 
m_{u,d} = P\ \wt{m}_{u,d}\ P\;,
\eeq
where $\wt{m}_{u,d} = m_{u,d}(\phi=\a=\j=0)$ and
\beq
P = \left(\begin{array}{ccc}
    e^{i(\phi - \a/2)}  & 0  & 0 \\
    0  & e^{i\a/2} & 0 \\
    0  &  0  & e^{i\j/2} 
    \end{array}\right) \;.
\eeq
It is then straightforward, but more tedious
than in the hermitean case (cf.~\cite{rr01}), 
to relate the phases of the mass matrix to those in the CKM matrix.
We obtain $\beta_{CKM}= -\chi - \omega$ and 
$\gamma_{CKM}= \pi + \chi - \Delta$, 
where $\Delta = \phi_u-\a_u - \phi_d + \a_d$ and $\chi$ is defined in ref.
\cite{rr01}. The angle $\omega$ is 
a function of the phases and the real parameters.
For $(\e_u)^2  \ll (\e_d)^2$ we have $\omega = \omega_d = \beta_d - (\a_d
+ \j_d)/2$. Data implies $\D \simeq  \pi/2$ with correspondingly 
smaller values  for $\chi$ and $\omega$.

Further information on the phases, in particular on relations between
phases in the up- and down-quark mass matrices can come from 
theoretical consistency conditions. In this connection it might be 
interesting that the
QCD $\Q$-parameter, which controls strong CP violation, is not renormalized
if the quark mass matrices satisfy the condition 
$\mbox{Im}\{\det{(m_u m_d)}\} = 0$ \cite{sw79}. This suggests the phase 
relation
\beq\label{cons}
2 \phi_u+\j_u + 2\phi_d+\j_d = n~\pi\;
\eeq   
with integer n. It would go beyond the purpose of this paper 
to discuss all the possible
solutions of this equation. We note, however, that if also  $\omega_u$ is
small and relation (\ref{special}) below holds, $\phi_u \simeq \j_u$ and
there are two solutions, $\phi_u \simeq \pm\pi/4$,
with $\alpha_u = 0$, depending on the value of n in eq. (\ref{cons}).

We do not know the structure of the Majorana mass matrix $M = h_N \VEV \Phi$.
It may be independent of the Higgs field, as in models with family
symmetries. In this case, one expects the same texture zeroes as
in the quark mass matrices,
\beq\label{Mtext}
M = \left(\begin{array}{ccc}
    0  & M_{12}  & 0 \\
    M_{12}  & M_{22} \; & M_{23} \\
    0  &  M_{23}  & M_{33} 
    \end{array}\right) \;,
\eeq
with $M_{12} \ll M_{22} \sim M_{23} \ll M_{33}$. $M$ is diagonalized by
a unitary matrix $U^{(N)}$,
\beq
U^{(N)\dg} M U^{(N)*} = \left(\begin{array}{ccc}
    M_1  & 0  & 0 \\
    0  & M_2 \; & 0 \\
    0  &  0  & M_3 
    \end{array}\right) \;.
\label{hevmix}
\eeq
Using the seesaw formula
we can now evaluate the light neutrino mass matrix. Since the choice of
the Majorana matrix $m_N$ fixes a basis for the right-handed neutrinos the
allowed phase redefinitions of the Dirac mass matrix $m_D$ are restricted.
In eq.~(\ref{qmass}) we have therefore kept the phases of all matrix elements. 

The $\n_\m$-$\n_\t$ mixing angle is known to be large. This leads us to
require $m_{\n_{i,j}}={\cal O}(1)$ for $i,j =2,3$. It is remarkable that this
determines the hierarchy of the heavy Majorana mass matrix to be\footnote{We
also note that this result is independent of the zeroes in the mass matrix
(\ref{qmass}) if its 1-3 element is smaller than $\e^3$, as required by
data.}
\beq
M_{12} : M_{22} : M_{33} = \e^5 : \e^4 : 1\;.
\eeq  
With $M_{33} \simeq M_3$, $M_{22} = \s \e^4 M_3$, 
$M_{23} = \z \e^4 M_3 \sim M_{22}$ and
$M_{12} = \e^5 M_3$, one obtains for masses and mixings to order 
${\cal O}(\e^4)$
\beq\label{Mhier}
M_1 \simeq - {\e^6\over \s}  M_3\;, \quad M_2 \simeq \s \e^4 M_3\;,
\eeq
\beq\label{Mmix}
U^{(N)}_{12} = - U^{(N)}_{21} = {\e\over \s}\;,\quad
U^{(N)}_{23} = {\cal O}(\e^4)\;, \quad
U^{(N)}_{13} = 0\;.
\eeq
Note, that $\s$ can always be chosen real whereas $\z$ is in general complex.
The inverse matrix reads, to leading order in $\e$,
\beq\label{itext}
M^{-1} = \left(\begin{array}{ccc}
    -\s  & \e  & -\z\e^3 \\
    \e  & 0  & 0 \\
    -\z\e^3  &  0  & \e^6 
    \end{array}\right)\ {1\over \e^6 M_3} \;.
\eeq
This yields for the light neutrino mass matrix
\beq\label{nmass1}
 m_{\n}  = - \left(\begin{array}{ccc}
    0  & \e e^{2i\phi}  & 0\\
    \e e^{2i\phi}   & -\s e^{2i\phi} + 2\r e^{i\phi}   & \h e^{i\phi}  \\
    0  & \h e^{i\phi}   & \; e^{2i\j}
    \end{array}\right)\ {v_1^2 \over M_3}  \;.
\eeq
The complex parameter
$\z$ does not enter because of the hierarchy.
Since, as required, all elements of the 2-3 submatrix are ${\cal O}(1)$, 
the mixing angle $\Q_{23}$ is naturally large. A large mixing 
angle $\Q_{12}$ may occur in case of a small determinant of the
2-3 submatrix \cite{vis98},
\beq\label{det}
(-\s + 2\r e^{-i\phi}) e^{2i\j} - \h^2 \equiv \d e^{2i\g} 
 = {\cal O}(\e)\;.
\eeq
Such a condition can be fullfilled without fine tuning if $\s, \r, 
\h = {\cal O}(1)$. It implies relations between the moduli as well as the 
phases of $\r$ and $\h$. In the special case of a somewhat smaller mass of the second
heavy neutrino, i.e., $|\s| < |\r|$, the condition (\ref{det}) becomes
\beq\label{special}
\j - \b \simeq {1\over 2} (\phi - \a)\;, \quad |\h|^2 \simeq 2 |\r|\;.
\eeq

The mass matrix $m_\n$ can again be diagonalized by a unitary matrix 
$U^{(\n)}$,
\beq\label{ndiag}
U^{(\n)\dg} m_{\n} U^{(\n)*}  = - \left(\begin{array}{ccc}
    m_1  & 0  & 0\\
    0   &  m_2   & 0  \\
    0  & 0    & m_3
    \end{array}\right) \;.
\eeq
A straightforward calculation yields ($s_{ij} = \sin{\Q_{ij}}$,
$c_{ij} = \cos{\Q_{ij}}$, $\xi=\e/(1+|\h|^2)$),
\beq\label{nmix}
U^{(\n)}  = \left(\begin{array}{ccc}
 c_{12} e^{i(\phi-\b+\j-\g)}  & s_{12} e^{i(\phi-\b+\j-\g)}  & 
             \xi s_{23} e^{i(\phi-\b+\j)}\\
 - c_{23}s_{12} e^{i(\phi+\b-\j+\g)} & c_{23}c_{12} e^{i(\phi+\b-\j+\g)}  & 
              s_{23} e^{i(\phi+\b-\j)}  \\
  s_{23}s_{12} e^{i(\g+\j)} & -s_{23}c_{12} e^{i(\g+\j)}   & c_{23} e^{i\j}
    \end{array}\right) \;,
\eeq
with the mixing angles,
\beq
\tan{2\Q_{23}} \simeq {2|\h|\over 1-|\h|^2}\;, \quad
\tan{2\Q_{12}} \simeq 2 \sqrt{1+|\h|^2} {\e \over \d}\;.
\eeq
Note, that the 1-3 element of the mixing matrix is small, 
$U^{(\n)}_{13} = {\cal O}(\e)$. The masses of the light neutrinos are
\bea
m_1 &\simeq& - {\e \over (1+|\h|^2)^{3/2}} 
        {(1-\cos{2\Q_{12}})\over \sin{2\Q_{12}}}\ m_3 \;,\\
m_2 &\simeq& {\e \over (1+|\h|^2)^{3/2}} 
        {(1+\cos{2\Q_{12}})\over \sin{2\Q_{12}}}\ m_3 \;,\\
m_3 &\simeq& (1+|\h|^2)\ {v_1^2\over M_3}\;.
\eea 
This corresponds to the weak hierarchy,
\beq
m_1 : m_2 : m_3 = \e : \e : 1 \;,
\eeq 
with $m_2^2 \sim m_1^2 \sim \D m_{21}^2 = m_2^2-m_1^2 \sim \e^2$. Since
$\e \sim 0.1$, this pattern is consistent with the LMA solution of the
solar neutrino problem, but not with the LOW solution.

We have obtained the large $\n_\m$-$\n_\t$ mixing as consequence of the 
required very large mass hierarchy (\ref{Mhier}) of the heavy Majorana 
neutrinos. The large $\n_e$-$\n_\m$ mixing follows from the
particular values of parameters ${\cal O}(1)$ for which we have not
found a particular reason. Hence, one expects two large mixing angles, but
single maximal or bi-maximal mixing would require strong fine tuning 
within our 
framework. On the other hand, a definite prediction is exactly one small 
matrix element, $U^{(\n)}_{13} = {\cal O}(\e)$. 

This pattern of neutrino mixings is a direct consequence of the hierarchy of the 
heavy 
Majorana masses and is independent of the off-diagonal elements of the 
mass matrix $M$. For instance, replacing the texture (\ref{Mtext}) by a
diagonal matrix, $M = \mbox{diag}(M_1,M_2,M_3)$,
yields the light neutrino mass matrix
\beq\label{nmass2}
m_{\n} = \left(\begin{array}{ccc}
    \e^2 e^{2i\phi}  & \r\e e^{i\phi} & \h\e e^{i\phi}\\
    \r\e e^{i\phi}  & \; a \; & \; b \\
    \h\e e^{i\phi}  &  \; b  & \; c
    \end{array}\right)\ {v_1^2\over M_3} \;.
\eeq
For the hierarchy,
\beq\label{hierM}
M_1 : M_2 : M_3 = \e^6 : \e^4 : 1 \;,
\eeq
the parameters $a$, $b$ and $c$ are again all ${\cal O}(1)$. 
The mass matrix (\ref{nmass2}) was previously obtained from  
a U(1) flavour symmetry \cite{sy98} where its structure is a consequence
of the U(1) charges of the lepton doublets and is
unrelated to the mass hierarchy of the heavy neutrinos.
This is in stark contrast to the SO(10) framework used here, where
the structure of $m_\n$ is intimately related to the hierarchy (\ref{hierM}).
Correspondingly, the assignment of U(1) charges is incompatible with the
SO(10) multiplet structure which may appear as an unsatisfactory feature 
of models with U(1) family symmetry.

In order to calculate various observables in neutrino physics we need
the leptonic mixing matrix 
\beq
U = U^{(e)\dg} U^{(\n)}\;,
\eeq
where $U^{(e)}$ is the charged lepton mixing matrix. In our framework we
expect $U^{(e)} \simeq V^{(d)}$, and also $V = V^{(u)\dg} V^{(d)}
\simeq V^{(d)}$ for the CKM matrix since $\e_u < \e_d$. This yields
for the leptonic mixing matrix 
\beq\label{lmix}
U \simeq  V^{\dg} U^{(\n)} \;.
\eeq
To leading order in the Cabibbo angle $\l \simeq 0.2$ we only need the
off-diagonal elements $V^{(d)}_{12} = \Bar{\l} = - V^{(d)*}_{21}$. Since
the matrix $m_d$ is complex, the Cabibbo angle is modified by phases,
$\Bar{\l} = \l\exp{\{i(\phi_d-\a_d)\}}$. The leptonic mixing matrix then
reads explicitly,
\beq
U = \left(\begin{array}{ccc} U_1 & U_2 & U_3\end{array}\right) \;,
\eeq
with the column vectors
\beq
U_1 = 
\left(\begin{array}{c} 
c_{12} e^{i(\phi-\b+\j-\g)} + \Bar{\l} c_{23}s_{12} e^{i(\phi+\b-\j+\g)}\\
-c_{23}s_{12} e^{i(\phi+\b-\j+\g)} + \Bar{\l}^* c_{12} e^{i(\phi-\b+\j-\g)}\\
s_{23}s_{12} e^{i(\g+\j)}\end{array}\right) \;,
\eeq
\beq
U_2 =
\left(\begin{array}{c}
s_{12} e^{i(\phi-\b+\j-\g)} - \Bar{\l} c_{23}c_{12} e^{i(\phi+\b-\j+\g)}\\ 
c_{23}c_{12} e^{i(\phi+\b-\j+\g)} + \Bar{\l}^* s_{12} e^{i(\phi-\b+\j-\g)}\\
-s_{23}c_{12}e^{i(\g+\j)}\end{array}\right) \;,
\eeq
\beq
U_3 =
\left(\begin{array}{c}
\xi s_{23} e^{i(\phi-\b+\j)} - \Bar{\l} s_{23} e^{i(\phi+\b-\j)} \\
s_{23} e^{i(\phi+\b-\j)} + \Bar{\l}^* \xi s_{23} e^{i(\phi-\b+\j)}\\
c_{23} e^{i\j}\end{array}\right) \;.
\eeq
Note, that all matrix elements are ${\cal O}(1)$ except $U_{13}$, where
we have counted the Cabibbo angle $\l = {\cal O}(\e)$. This matrix
element is predicted to be close to the experimental limit, 
\beq
|U_{13}| =  {\cal O}(\l,\e) \sim 0.1 \;.
\eeq

Next, we consider CP violation in neutrino oscillations. Obervable 
effects are  controlled by the Jarlskog parameter $J_l$ \cite{jar85}
($\wt{\e}_{ij} = \sum_{k=1}^3 \e_{ijk}$)
\beq
\mbox{Im}\{ U_{\a i}U_{\b j}U_{\a j}^*U_{\b i}^*\} 
= \wt{\e}_{\a\b}\wt{\e}_{ij} J_l\;,
\eeq
for which we find 
\beq
J_l \simeq  \l s_{12}c_{12}c_{23}s_{23}^2 \sin{(2(\b-\j+\g)+\phi_d-\a_d)}\;.
\eeq
In the case of a small mass difference $\D m_{12}^2$ the CP asymmetry 
$P(\n_\m\rightarrow \n_e) - P(\Bar{\n}_\m\rightarrow \Bar{\n}_e)$
is proportinal to $\d$ (cf.~(\ref{det})). Hence, the dependence of $J_l$ 
on the angle $\g$ is not surprising.

For large mixing, $c_{ij}\simeq s_{ij} \simeq 1/\sqrt{2}$, and
in the special case (\ref{special}) we find from the SO(10) phase relation
$\phi-\a = \phi_u -\a_u$ and $\phi_u -\a_u - \phi_d +\a_d = \Delta 
\simeq \pi/2$,
\beq
J_l \simeq {\l\over 4\sqrt{2}} \sin{\left(-{\pi\over 2} + 2\g\right)}\;.
\eeq
For small $\g$ this corresponds to maximal CP violation, but without a
deeper understanding of the fermion mass matrices this case is not
singled out.
Due to the large neutrino mixing angles, $J_l$ is much bigger than the
Jarlskog parameter in the quark sector, $J_q = {\cal O}(\l^6) \sim 10^{-5}$,
which may lead to observable effects at future neutrino factories \cite{blo00}.

According to the seesaw mechanism neutrinos are Majorana fermions. This
can be directly tested in neutrinoless double $\b$-decay. The decay
amplitude is proportional to the complex mass 
\bea
\VEV m &=& \sum_i U_{ei}^2 m_i 
= - (U U^{(\n)\dg} m_\n U^{(\n)*} U^T)_{ee} \simeq - (V^{(d)\dg} m_\n V^{(d)*})_{ee} \NO\\ 
&=& - {1 \over 1 + |\h|^2} \left(\l^2 |\h|^2 e^{2i(\phi_d -\a_d + \b + \phi- \j)}
                             - 2 \l \e e^{i(\phi_d -\a_d + 2 \phi)}\right) m_3\;. 
\eea
With $m_3 \simeq \sqrt{\D m^2_{atm}} \simeq 5\times 10^{-2}$~eV this yields 
$\VEV m \sim  10^{-3}$~eV, more than two orders of magnitude
below the present experimental upper bound \cite{hm99}. 

We now turn to  the cosmological matter-antimatter
asymmetry.  An attractive mechanism to generate it is leptogenesis \cite{fy86} 
which involves both, CP violation
and the Majorana nature of the neutrinos. This connection has already
been discussed in different contexts \cite{jpr01,bmx01}.
The baryon asymmetry is given by
\beq\label{basym}
Y_B = {n_B-n_{\Bar{B}}\over s} = \k c_S {\ve_1\over g_*}\;.
\eeq
Here $n_B$ and $s$ are baryon number and entropy densities, respectively.
$g_* \sim 100$ is the number of degrees of freedom in the plasma of the
early universe, $\ve_1$ is the CP asymmetry in the decay of the lightest
of the heavy Majorana neutrinos and $c_S$ is the conversion factor from
lepton asymmetry to baryon asymmetry due to sphaleron processes. For
three quark-lepton generations and two Higgs doublets one has
$c_S = - 8/15$ \cite{ht90}. The effects of washout 
processes are accounted for by $\k <1$.

It is convenient to express the CP asymmetry directly in terms of the
light neutrino mass matrix. In a flavour diagonal basis for the heavy
neutrinos one has \cite{bf00}
\beq
\ve_1 \simeq - {3\over 16\pi} {M_1\over (h_\n^\dg h_\n)_{11}}
 \mbox{Im}\left(h_\n^\dg h_\n {1\over M} h_\n^T h_\n^*\right)_{11}\;.
\eeq
In an arbitrary basis for light and heavy neutrinos this can be rewritten as
\beq
\ve_1 \simeq {3\over 16\pi} \mbox{sign}(M_1) 
{\mbox{Im}\left(U^{(N)T} m_D^\dg m_\n m_D^* U^{(N)}\right)_{11} 
 \over v_1^2 \wt{m}_1}\;,
\eeq
where $U^{(N)}$ is the heavy neutrino mixing matrix defined in 
eq. (\ref{hevmix}). The effective neutrino mass
$\wt{m}_1 = \left(U^{(N)T} m_D^\dg m_DU^{(N)*}\right)_{11}/ |M_1|$
is a sensitive parameter for successful leptogenesis \cite{plu98}.
From eqs.~(\ref{qmass}), (\ref{Mmix}) and (\ref{nmass1}) one then obtains
\beq 
\ve_1 \simeq {3\over 16\pi} \e^6 {|\h|^2 \over \s}
{\left(|\r|^2 \sin{(2(\phi-\a+\b-\j))}+2 |\r| \sin{(\phi-\a)} 
+ \sin{(2(\j-\b))}\right) \over \s^2 + |\h|^2 +|\r|^2 - 
2|\r|\s \cos(\phi - \a)}\;. 
\eeq  
In the special case (\ref{special}) this expression simplifies to
\beq
\ve_1 \simeq {3\over {16\pi}} \e^6 {|\h|^2\over \s}  
         {(1+|\r|)^2 \over |\h|^2 +|\r|^2}\sin(\phi-\a) \;.
\eeq  

With $\e \sim 0.1$ one has $\ve_1 \sim 10^{-7}$ ,
$|M_1| \simeq (\e^6/|\s|) (1+|\h|^2) v_1^2/m_3 \sim 10^9$~GeV and 
$\wt{m}_1 \sim (|\h|^2+|\r|^2)/(\s(1+|\h|^2)) m_3 \sim 10^{-2}$~eV. 
The baryon asymmetry is then given by
\beq\label{asym}
Y_B \sim  - \k~\mbox{sign}(\s)~\sin{(\phi-\a)} \times 10^{-9}\;.
\eeq
From the SO(10) symmetry one obtains $\phi-\a = \phi_u-\a_u$. 
According to the qualitative discussion below eq. (\ref{cons}), there are
two solutions with $\alpha_u \simeq 0$, $\phi_u \simeq \pi/4$ and
$\phi_u \simeq -\pi/4$. Thus, depending on the sign of $\s$, there
is always a positive baryon asymmetry, in agreement with observation. 
Without further assumptions, the values of $\phi_u-\a_u$ and $\s$ cannot be 
fixed.

The parameters $\ve_1$, $M_1$ and $\wt{m}_1$ are rather similar to those 
considered previously in a leptogenesis scenario \cite{bp96,by99} with 
hierarchical heavy Majorana neutrinos and with $B-L$ broken at the GUT scale.  
We therefore expect that a solution of the full Boltzmann equations will
yield a baryon asymmetry which is consistent with the observed asymmetry
$Y_B \simeq (0.6 - 1)\times 10^{-10}$.

In summary, we have considered the consequences of large neutrino mixing,
as indicated by data, in connection with SO(10) symmetry and the
seesaw mechanism. This determines uniquely the hierarchy of the 
heavy Majorana neutrino masses. 
The resulting light neutrino mass hierarchy
is consistent with the LMA solution of the solar neutrino problem
but incompatible with the LOW solution. Furthermore, the $U_{13}$ element
of the leptonic mixing matrix is predicted to be 
$U_{13} = {\cal O}(\l,\e) \sim 0.1$. CP violation in neutrino 
oscillations may be maximal, and the correct order of magnitude for the baryon 
asymmetry  is obtained. In the case of a very large
heavy neutrino mass hierarchy, i.e. $ ({M_2 / M_3}) < ({m_{charm} 
/m_{top}})^2$, the baryon asymmetry is more closely related to the CP 
violating phases
in the quark sector and the correct sign can be obtained.
However, a complete determination of the magnitude and the 
relative sign of both
CP violating observables requires a deeper understanding of the quark
and lepton mass matrices.

\section*{Acknowledgements}

We thank M. Pl\"umacher for pointing out a sign error and very helpful 
comments on the text. We also thank G. Colangelo for useful remarks,
and R. Rosenfeld and J. Rosner for a correspondence.

\end{document}